\documentclass[twocolumn,english]{revtex4-1}
\usepackage[T1]{fontenc}
\usepackage[utf8]{inputenc}
\usepackage{color}
\usepackage{bm}
\usepackage{amsmath}
\usepackage{graphicx}

\makeatletter

\newcommand{\lyxdot}{.}

\makeatother

\usepackage{babel}
\begin{document}

\title{Study of Active Brownian Particle Diffusion in Polymer Solutions}

\author{Yunfei Du, Zhonghuai Hou{*}}

\address{Department of Chemical Physics and Hefei National Laboratory for
Physical Sciences at Microscales, iChEM, University of Science and
Technology of China, Hefei, Anhui 230026, China}
\begin{abstract}
The diffusion behavior of an active Brownian particle (ABP) in polymer
solutions is studied using Langevin dynamics simulations. We find
that the long time diffusion coefficient $D$ can show a non-monotonic
dependence on the particle size $R$ if the active force $F_{a}$
is large enough, wherein a bigger particle would diffuse faster than
a smaller one which is quite counterintuitive. By analyzing the short
time dynamics in comparison to the passive one, we find that such
non-trivial dependence results from the competition between persistence
motion of the ABP and the length-scale dependent effective viscosity
that the particle experienced in the polymer solution. \textcolor{black}{We
have also introduced an effective viscosity $\eta_{\text{eff}}$ experienced
by the ABP phenomenologically. Such an active $\eta_{\text{eff}}$
is found to be larger than a passive one and strongly depends on $R$
and $F_{a}$}\textcolor{magenta}{.} In addition, we find that the
dependence of $D$ on propelling force $F_{a}$ presents a well scaling
form at a fixed $R$ and the scaling factor changes non-monotonically
with $R$. Such results demonstrate that active issue plays rather
subtle roles on the diffusion of nano-particle in complex solutions.
\end{abstract}
\maketitle

\section{INTRODUCTION}

Transport properties of macromolecule are of great importance in various
field\textcolor{black}{s} including biophysics\citep{wang2012disordered,sigalov2010protein,pederson2000diffusional,cluzel2000ultrasensitive},
material \citep{struntz2015multiplexed,gersappe2002molecular,zhou2007keys,shah2005effect},
medicine\citep{rohilla2016herbal,huo2014redox}, \textcolor{black}{and
so on}. One of the research interests in recent years is the diffusion
of macromolecule (protein or nano-particle) in complex solution system\textcolor{black}{s}.
Especially in living cells, a variety of structural and functional
proteins\textcolor{black}{{} are} immersed in crowded cytoplasmic environments,
involved in diverse biochemical processes such as enzyme reactions\citep{berry2002monte},
signal transmission\citep{wang2012disordered}, gene transcription
and self-assembly of supramolecular\citep{macnab2000action}.\textcolor{black}{{}
Experimentally, one would usually focus on nano-particle (NP) in complex
environment such as polymer solution. }And study of diffusive behavior
of NP can provide important information about the local structure
, viscoelastic properties and also the crowding effect of polymer
liquids\citep{lu2002probe,waigh2005microrheology,cicuta2007microrheology}.

In the past decades, diffusion of NPs in polymer solutions has received
a lot of attention both experimentally \citep{Koh126143,Che16106101,Pry162145,Lee17406,Pol151169}
and theoretically \citep{Tui08254,Gan06221102,Yam14152,Don15024903,Fen1610114,Ego11084903,Yam11224902}.
In experiments, fluctuation correlation spectroscopy (FCS) \citep{Oma098449,Hol099025,Mic071595,Gra09021903},
dynamic light scattering (DLS) \citep{Hol099025,Koe04021804}, and
capillary viscosimetry are general tools to investigate the diffusion
of a NP in complex fluids. It is found that as the particle radius
$R$ decreases to nanoscale, the diffusion coefficient $D$ would
increase exponentially with $R$ and violates the Stokes-Einstein
\textcolor{black}{(SE)}\textcolor{magenta}{{} }relation apparently\citep{ye1998transport,schachman1952viscosity,tuteja2007breakdown,kohli2012diffusion,wang2010effects}.
Phillies $et$ $al$, based on the analysis of a great deal of experiment\textcolor{black}{al
data}, proposed an empirical formula $D=D_{0}\exp(-\alpha c^{v})$,
where $D_{0}$ is the diffusion coefficient in a purely background
\textcolor{black}{solvent}, $c$ is the concentration of polymer solution
and $\alpha$, $v$ are fitting \textcolor{black}{parameters} relevant
to a specific system\citep{phillies1986universal}. Although this
stretched exponential form matches the experimental data very well,
the underlying physical meaning of these two parameters are \textcolor{black}{quite}
blurry\citep{kalwarczyk2015motion}. Recently, Holyst $et$. $al$
proposed a length-scale dependent viscosity theory\citep{holyst2009scaling,kalwarczyk2011comparative}.
They argued that the diffusion of a NP in polymer solution was characterized
by at least three length scales: the particle size $R$, the polymer
hydrodynamic radius $R_{h}$ and the correlation length $\xi$ of
the solution. Concretely, the formula reads $D/D_{0}\sim\exp\left[b\left(R_{\text{eff}}/\xi\right)^{a}\right]$,
where $R_{\text{eff}}=\sqrt{R^{2}R_{h}^{2}/(R^{2}+R_{h}^{2})}$ denotes
an effective size, $a$ and $b$ are fitting parameters. \textcolor{black}{If
$R$ is small with respect to $\xi$, the particle motion experiences
the local viscosity which is smaller than the macro-viscosity by order
of magnitude. While if $R$ is much larger than $R_{h}$ or $R_{g}$,
the particle motion would no more be affected by the local structures
of polymer and touches the macro viscosity finally, and obeys the
Stokes-Einstein relation automatically. }Also, there are some other
interesting models in this field such as the hopping model\citep{cai2015hopping},
walking confined diffusion model\citep{ochab2011scale} and depletion
model\citep{tuinier2008scaling}. Liu $et$ $al$ used a MD simulation
to study the diffusion of a NP in polymer melt \citep{liu2008molecular}and
the core results are similar to Holyst's work, where $R_{g}$ could
be the boundary of NP size to experience the local viscosity to macro
viscosity. Note that, it is hard to take a simulation work on this
issue especially the polymer solution, which on one side the solvent
accounts for a large proportion and costs a huge computational resource,
and on other side a reasonable diffusion coefficient needs so much
long time to evolve the system . To our best of knowledge, only in
the recent two years, Li $et$ $al$ \citep{Li2016Diffusion,chen2017effect}and
Pryamitsyn $et$ $al$\citep{pryamitsyn2016noncontinuum} had studied
this issue in a simulation way, using Multiparticle Collision Dynamic
(MPCD) and Dissipative Particle Dynamic (DPD) method respectively.

Most of the present works focus on the diffusion of passive NP. But
note that, in real biological system, especially in a living cell,
the active proteins widely exist. Typical active proteins include
motor molecule\citep{zheng2000prestin,duan2016ubiquitin}, microtubule
or active filament\citep{sumino2012large,loose2014bacterial}. By
consuming the ATP, they get a propel force that extremely enhance
the transport efficiency in various biochemical process. Actually,
the dynamics behavior of active matter has gained much attention in
recent years. Instead of the living species such as bacteria\citep{sokolov2007concentration},
spermatozoa\citep{woolley2003motility,riedel2005self} and the micro
protein in cell as mentioned above, there are also much artificial
objects capable of self-propulsion such as Janus particles\citep{howse2007self},
chiral particles\citep{ghosh2009controlled}, and vesicles\citep{joseph2016active}.
And a wealth of new non-equilibrium phenomena have been reported,
including phase separation\citep{fily2012athermal}, active turbulence\citep{dunkel2013fluid},
and active swarming\citep{cohen2014emergent}, both experimentally
and theoretically.

Recently, much interests arised on the dynamic behaviors of an active
particle in complex environment\citep{shen2011undulatory,gagnon2014undulatory,thomases2014mechanisms,patteson2016active}.
And a number of studies indicated that there exists a two-way coupling
between the active matter and the ambient environment, which the motion
of active suspensions can alter the local property of its environment,
while simultaneously the complex fluid rheology can modify the dynamics
of the active matters\citep{patteson2016active}. For instance, Patteson
$et$ $al$ reported an experiment on the diffusion of an Escherichia
coli in polymeric solution\citep{patteson2015running}. They found
that the translational diffusion of cell is enhanced and the rotational
diffusion is sharply declined respected to the diffusion behavior
in water-like fluid, due to the complicated interaction with the polymers
in solution.\textcolor{black}{{} It was also found that activity has
a fascinating effect on the viscosity of active suspensions, even
sometimes leading to a ``vanishing'' viscosity phenomenon in bacterial
suspensions\citep{lopez2015turning}. Even for a single swimmer, there
is no universal answer to whether mobility is enhanced or hindered
by fluid elasticity\citep{patteson2016active}. Despite lot of interesting
progresses made so far, there still remains many open questions to
be answered, even some fundamental ones. For instance, how would the
long time diffusion coefficient of an active particle depends on its
size in a complex fluid, although being a quite straightforward question,
has not been systematically studied yet. }

\textcolor{black}{In the present work, we have addressed such a topic
by investigating the diffusion dynamics of an active Brownian particle
(ABP) in polymer solutions, as depicted schematically in Fig.1. The
ABP is modeled by a spherical particle subjected to an active force
along the direction denoted by $\mathbf{n}$, which changes randomly
with time. Three-dimensional Langevin simulations are performed to
calculate the long time diffusion coefficient $D$ of the ABP as a
function of the particle size $R$, for a variety of different active
force $F_{a}$ as well as polymer concentration $\phi$. Very interestingly,
we find that $D$ shows a non-monotonic dependence on $R$ if the
active force $F_{a}$ is large enough: $D$ first increases with the
particle size $R$, reaches a maximum value at an optimal particle
size $R_{\text{opt}}$, after which it decreases monotonically. The
optimal value $R_{\text{opt}}$ moves to a larger value with the increment
of active force $F_{a}$ and to a smaller value with increasing polymer
concentration. Further analysis of the time dynamics of the mean-square
displacement (MSD) indicates that it is the competition between the
persistence motion of the particle, which is the reason for superdiffusion,
and the cage effect of the polymer solution, which leads to the subdiffusion
behavior, that causes the optimal size effect for long time diffusion.
We have also introduced an effective viscosity experienced by the
ABP by introducing a phenomenological model describing the ABP moving
in a fluid with effective viscosity $\eta_{\text{eff}}^{a}$. Such
an effective viscosity is found to be larger than that experienced
by a passive particle, and it shows strong dependency on the particle
size $R$ as well as the active force amplitude $F_{a}$. In addition,
we have found that $D$ shows a power law dependence on the active
force $F_{a}$, i.e., $D\sim F_{a}^{\alpha}$, for a fixed particle
size. More interestingly, the exponent $\alpha$ also shows a non-monotonic
dependence on the particle size $R$: $\alpha<2.0$ for small $R$,
then it increases with $R$ to a maximum value $\alpha\sim2.5$ for
an intermediate particle size, and finally approaches 2.0 in the large
size limit. Our findings demonstrate that interplay between particle
activity and local structure in complex solution may lead to interesting
dynamics of the ABP.}

\textcolor{black}{}
\begin{figure}
\begin{centering}
\textcolor{black}{\includegraphics[bb=0bp 0bp 818bp 595bp,clip,width=0.8\columnwidth]{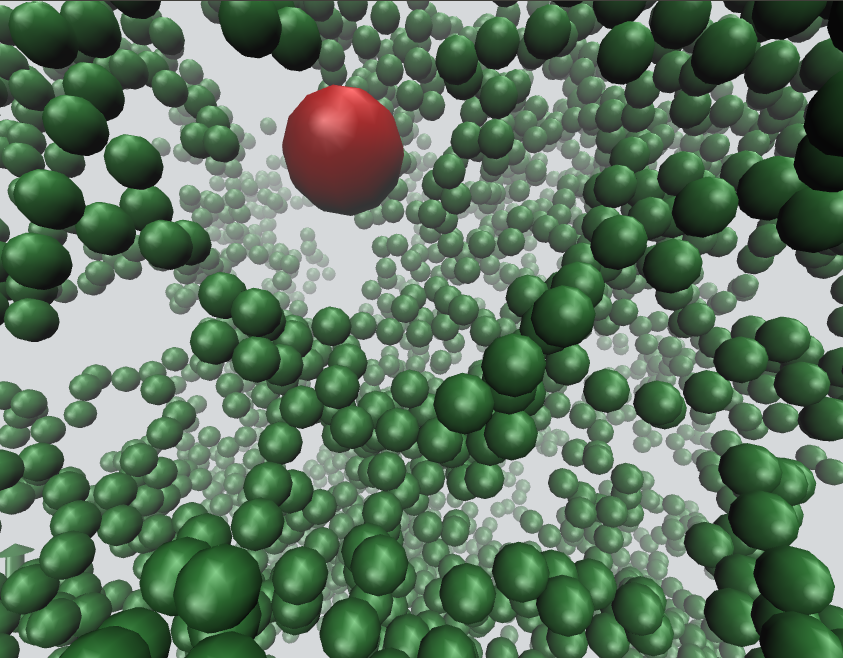}}
\par\end{centering}

\textcolor{black}{\caption{The illustration of a ABP immerses in polymer solution, which the
big red particle represents ABP and the green particle around represents
polymer bead. }
}
\end{figure}

\textcolor{black}{The paper is organized as follows. In section \ref{sec:Simulation-method}
, we describe the model and si}mulation method. Results and discussion
are presented in section \ref{sec:Results} followed by conclusion
in section \ref{sec:Conclusion} .

\section{\label{sec:Simulation-method}Simulation method}

\textcolor{black}{As shown in Fig.1, we consider a three dimensional
system containing a single ABP of radius $R$ in polymer solution.
The polymers are modeled as bead-spring chains, each consisting of
$N$ beads with diameter $\sigma$. All non-bonded (excluded volume)
interactions between the beads are modeled by purely repulsive Weeks-Chandler-Andersen
(WCA) potentials\citep{weeks1971role}:}

\textcolor{black}{
\begin{equation}
U_{\text{BB}}(r_{ij})=\begin{cases}
4\epsilon\left[(\frac{\sigma}{r_{ij}})^{12}-(\frac{\sigma}{r_{ij}})^{6}+\frac{1}{4}\right] & \qquad r_{ij}<\sqrt[6]{2}\sigma\\
0 & \qquad r_{ij}\geq\sqrt[6]{2}\sigma
\end{cases}\label{eq:1}
\end{equation}
where $r_{ij}=\left|\mathbf{r}_{i}-\mathbf{r}_{j}\right|$ denotes
the distance between the two beads $i$ and $j$ (with position vectors
given by $\mathbf{r}_{i}$ and $\mathbf{r}_{j}$, respectively), $\epsilon$
represents strength of the WCA potential. The bond interaction between
two nearby beads is modeled by the FENE (finite extensible nonlinear
elastic) potential\citep{kremer1990dynamics}:}

\textcolor{black}{
\begin{equation}
U_{\text{FENE}}(r_{ij})=-\frac{1}{2}\epsilon_{F}r_{F}^{2}\ln\left[1-(\frac{r_{ij}}{r_{F}})^{2}\right]\label{eq:2}
\end{equation}
where $\epsilon_{F}$ is the interaction strength and $r_{F}=2.0\sigma$
denotes the upper bound of $r_{ij}$. }

\textcolor{black}{The interactions between the NP and polymer beads
are also described by the truncated WCA potential which is offset
by the interaction range $R_{ev}=R-\sigma/2$:}

\textcolor{black}{
\begin{equation}
U_{BN}(r_{jn})=4\epsilon\left[(\frac{\sigma}{r_{jn}-R_{ev}})^{12}-(\frac{\sigma}{r_{jn}-R_{ev}})^{6}+\frac{1}{4}\right]\label{eq:3}
\end{equation}
for $R_{ev}<r_{jn}<R_{ev}+2^{1/6}\sigma$, where $r_{jn}$ denotes
the distance between bead $j$ and the nano-particle with position
given by $\mathbf{r}_{n}$. While for $r_{jn}\leq R_{ev}$, }$U_{BN}(r_{jn})=\infty$\textcolor{black}{{}
and for $r_{jn}\geq R_{ev}+2^{1/6}\sigma$, }$U_{BN}(r_{jn})=0$.

\textcolor{black}{The dynamics of the polymer beads are described
by the following Langevin equations(ignoring hydrodynamic interactions):}

\textcolor{black}{
\begin{equation}
m_{\text{B}}\frac{d^{2}\mathbf{r}_{j}}{dt^{2}}=-\gamma_{\text{B}}\frac{d\mathbf{r}_{j}}{dt}-\nabla_{\mathbf{r}_{j}}U'+\sqrt{2k_{\text{B}}T\gamma_{\text{B}}}\bm{\xi}_{j}(t)
\end{equation}
where $U'=\sum_{i\ne j}U_{\text{BB}}\left(r_{ij}\right)+U_{\text{BN}}\left(r_{jn}\right)$
, $m_{B}$ is the bead mass and $\gamma_{B}$ is the friction coefficient
of the bead in the background pure solvent, $\bm{\xi}_{j}\left(t\right)$
denotes independent Gaussian white noises with zero means and unit
variances, i.e., $\left\langle \bm{\xi}_{j}\left(t\right)\right\rangle =0$,
$\left\langle \bm{\xi}_{i}\left(t\right)\bm{\xi}_{j}\left(t'\right)=2\delta_{ij}\mathbf{I}\delta\left(t-t'\right)\right\rangle $
where $\mathbf{I}$ is the unit tensor. }

\textcolor{black}{The dynamics of the ABP is given by}

\textcolor{black}{
\begin{align}
m_{\text{N}}\frac{d^{2}\mathbf{r}_{n}}{dt^{2}} & =-\gamma_{\text{N}}\frac{d\mathbf{r}_{n}}{dt}+F_{a}\textbf{n}-\nabla_{\mathbf{r}_{n}}\left[\sum_{j}U_{\text{BN}}\left(r_{jn}\right)\right]+\bm{\xi}_{n}(t)\\
\frac{d\mathbf{n}}{dt} & =\bm{\eta}\left(t\right)\times\mathbf{n}
\end{align}
where $m_{N}$ is the mass of the ABP and $\gamma_{N}$ is the friction
coefficient of the ABP in the pure solvent. $\mathbf{r}_{n}$ is the
position vector of the ABP, $r_{jn}=\left|\mathbf{r}_{j}-\mathbf{r}_{n}\right|$
is the distance between polymer bead $j$ and the ABP. $F_{a}$ represents
the amplitude of active force with orientation specified by the unit
vector $\mathbf{n}$. $\bm{\xi}_{n}\left(t\right)$ is also a Gaussian
white noise vector with $\left\langle \xi_{n}\left(t\right)=0\right\rangle $
and $\left\langle \bm{\xi}_{n}\left(t\right)\bm{\xi}_{n}\left(t'\right)\right\rangle =2D_{t}\mathbf{I}\delta\left(t-t'\right)$,
where $D_{t}=k_{B}T/\gamma_{N}$ is the (short time) translational
diffusion coefficient. The stochastic vector $\bm{\eta}\left(t\right)$
is also Gaussian distributed with zero mean and has time correlations
given by $\left\langle \bm{\eta}\left(t\right)\bm{\eta}\left(t'\right)\right\rangle =2D_{r}\mathbf{I}\delta\left(t-t'\right)$,
where $D_{r}$ denotes the rotational diffusion coefficient. Since
we consider a spherical particle here, $D_{t}$ is related to $D_{r}$
via $D_{r}=3D_{t}/\left(2R\right)^{2}$.}

\textcolor{black}{All simulations were performed in a cubic box with
a $25\sigma$ edge length with periodic boundary conditions in all
directions. A value $\epsilon=k_{B}T$ was used for all particle interactions,
where $k_{\text{B}}$ is Boltzmann's constant and $T$ temperature.
The polymers were modeled using $N=64$ beads and the parameters for
bond-interactions are $k_{\text{F}}=10k_{\text{B}}T\sigma^{-2}$ and
$r_{\text{F}}=2.0\sigma$. If not otherwise specified, we considered
a system containing 72 polymer chains, corresponding to a bead-number
concentration $\phi\simeq0.3$. For other polymer concentrations,
we simply varied the number of polymer chains. We assumed equal densities
of the ABP and polymer bead, thus $m_{\text{N}}=8R^{3}/\sigma^{3}$$m_{\text{B}}$.
Since the friction coefficient $\gamma$ is proportional to $6\pi\eta_{0}R$
for a spherical particle of radius $R$ in the pure solvent, where
$\eta_{0}$ is the zero-shear viscosity of the pure solvent, we have
$\gamma_{\text{N}}=\left(2R/\sigma\right)\gamma_{\text{B}}$. We set
$m_{\text{B}}=1$, $k_{\text{B}}T=1$, $\sigma=1$ for dimensionless
units and then fixed $\gamma_{\text{B}}=1$. The remained variable
parameters are the active force amplitude $F_{a}$ and the particle
radius $R$. Velocity-Verlet algorithm was used to simulate the dynamic
equations with a time step $\Delta t=0.01$. All the reported data
below were obtained after averaging over 20 independent runs with
long enough time. }

\section{\label{sec:Results}Results And Discussion}

\subsection{Optimal Size for Active Particle Diffusion}

\begin{figure}
\begin{centering}
\includegraphics[width=0.8\columnwidth]{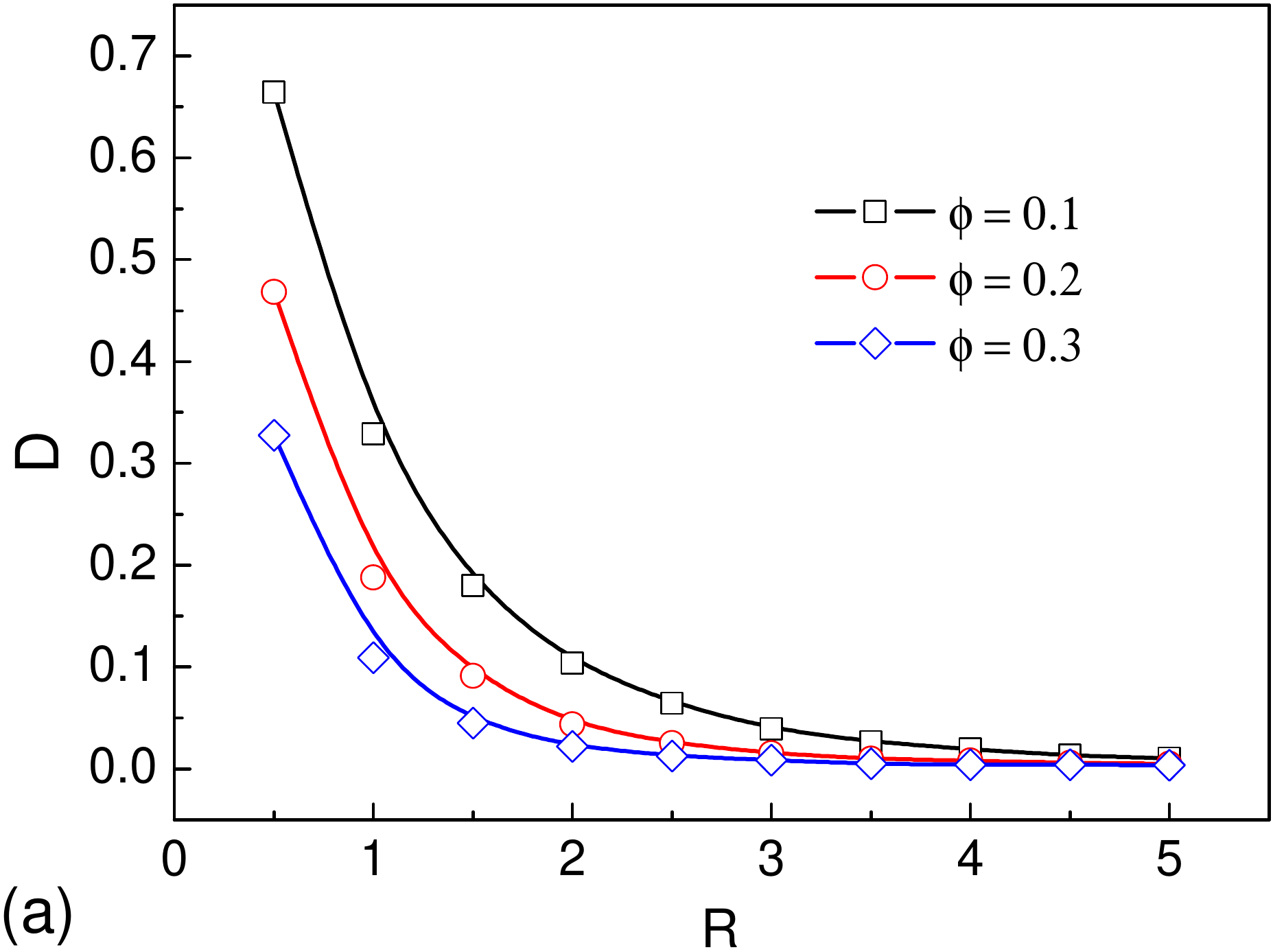}
\par\end{centering}

\begin{centering}
\includegraphics[width=0.8\columnwidth]{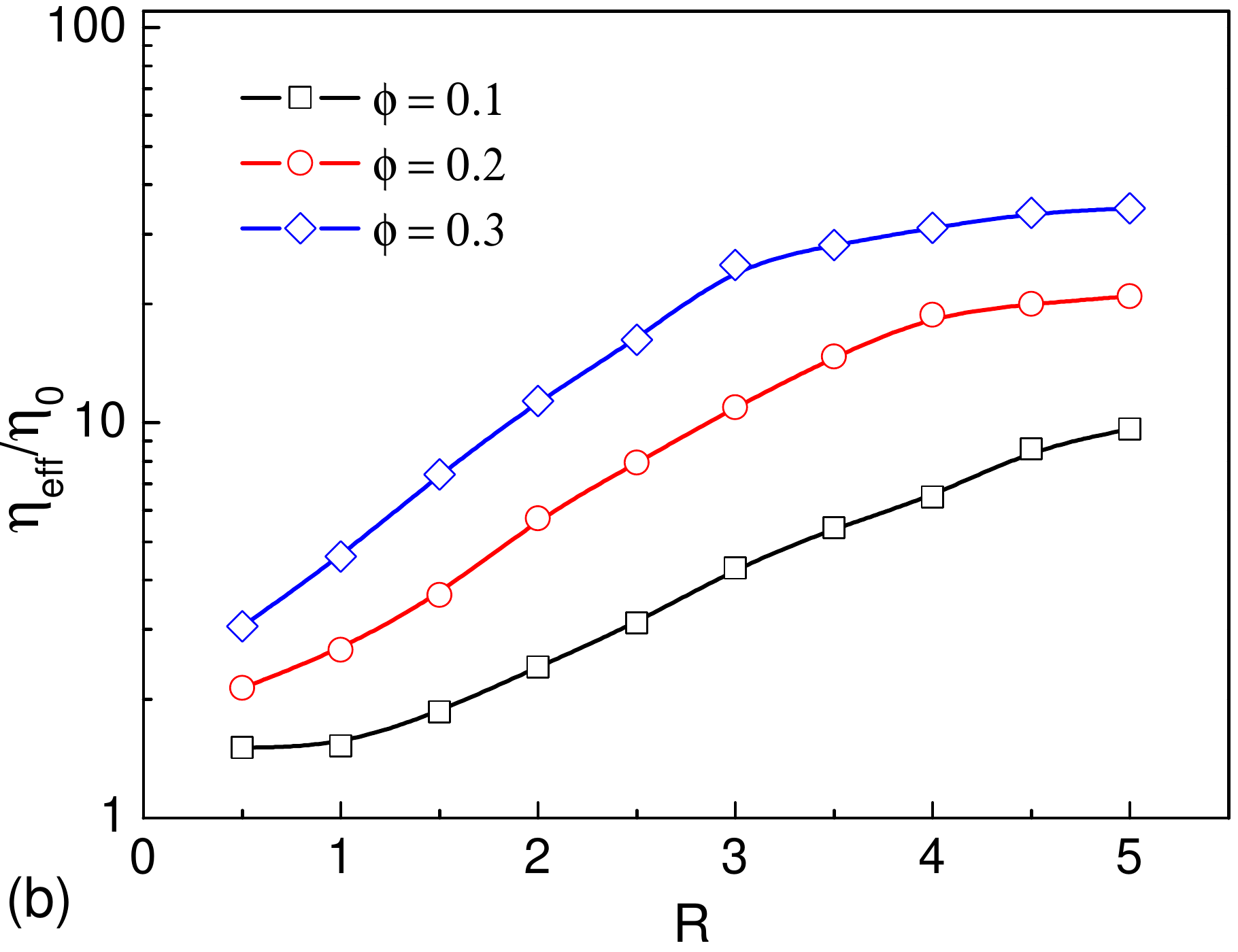}
\par\end{centering}

\caption{\textcolor{black}{The diffusion coefficient $D$ (a) and the effective
viscosity $\eta_{\text{eff}}$ (b) of a passive NP with respect to
different size $R$, under the concentration of polymer $\phi=0.1,$
0.2 and 0.3, respectively. Note that $\eta_{\text{eff}}$ is reduced
by the viscosity of pure solution $\eta_{0}$. The solid lines are
drawn to guide the eyes.}\textcolor{magenta}{{} }\label{fig:D_passive}}
\end{figure}

\textcolor{black}{In the present paper, we are mainly interested in
how the activity would influence the NP diffusion behavior in the
polymer solution. For comparison, we first investigate the diffusion
behavior of a passive particle in the system set above. The long time
diffusion coefficient $D$ is calculated via
\[
D=\ensuremath{\lim\limits _{t\to\infty}\frac{1}{6t}\langle\Delta r_{n}^{2}(t)\rangle}
\]
where $\langle\Delta r_{n}^{2}(t)\rangle=\langle|\textbf{r}_{n}(t)-\textbf{r}_{n}(0)|^{2}\rangle$
is the mean square displacement (MSD) of the NP with $\textbf{r}_{n}(t)$
being the particle position at time $t$. As already mentioned in
the introduction section, passive particle diffusion in polymer solutions
may strongly deviates the SE relation. In particular, the diffusion
coefficient $D$ can be described by scaling relations involving the
correlation length of the polymer solution and an effective size.
Note that the system parameters used in our present work may not fit
well the experimental conditions. In Fig.\ref{fig:D_passive}(a),
the diffusion coefficients $D$ of a passive NP as functions of the
radius $R$ for several fixed values of polymer concentration $\phi=0.1,\:0.2\text{ and }0.3$
are presented. Clearly, $D$ decreases monotonically with increasing
of $R$ as well as the polymer concentration $\phi$ as expected.
We can introduce an effective (so-called) nano-viscosity $\eta_{\text{eff}}$
experienced by the NP via the standard SE relation $D=k_{B}T/6\pi\eta_{\text{eff}}R$.
In the large particle size limit, $\eta_{\text{eff}}$ would be the
macroscopic zero-shear viscosity $\eta_{macro}$ of the polymer solution.
For small particle size, however, $\eta_{\text{eff}}$ is much smaller
than $\eta_{macro}$ leading to large deviations from the SE relation.
In Fig.\ref{fig:D_passive}(b), the nano-viscosity as a function of
the NP size $R$ is shown for different concentration $\phi$. $\eta_{\text{eff}}$
firstly increases fiercely with $R$ until it finally reaches the
macroscopic value $\eta_{macro}$ for large particle sizes. It suffices
to reach} $\eta_{macro}$ for a NP with size to be just a few times
of that of a polymer bead and NP in a more concentrated polymer solution
can reach $\eta_{macro}$ at a smaller $R$.

\begin{figure}
\begin{centering}
\includegraphics[width=0.8\columnwidth]{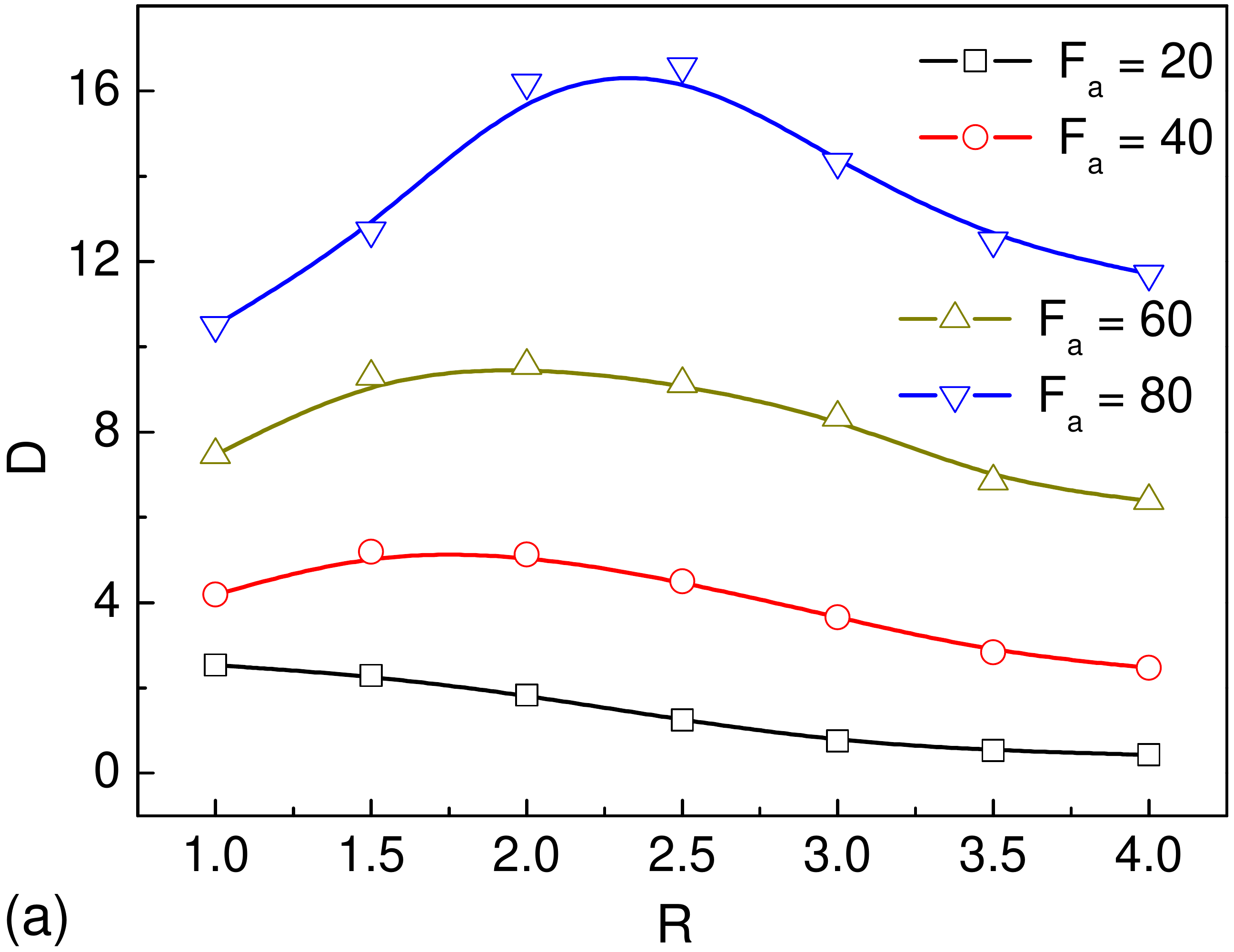}
\par\end{centering}

\begin{centering}
\includegraphics[width=0.8\columnwidth]{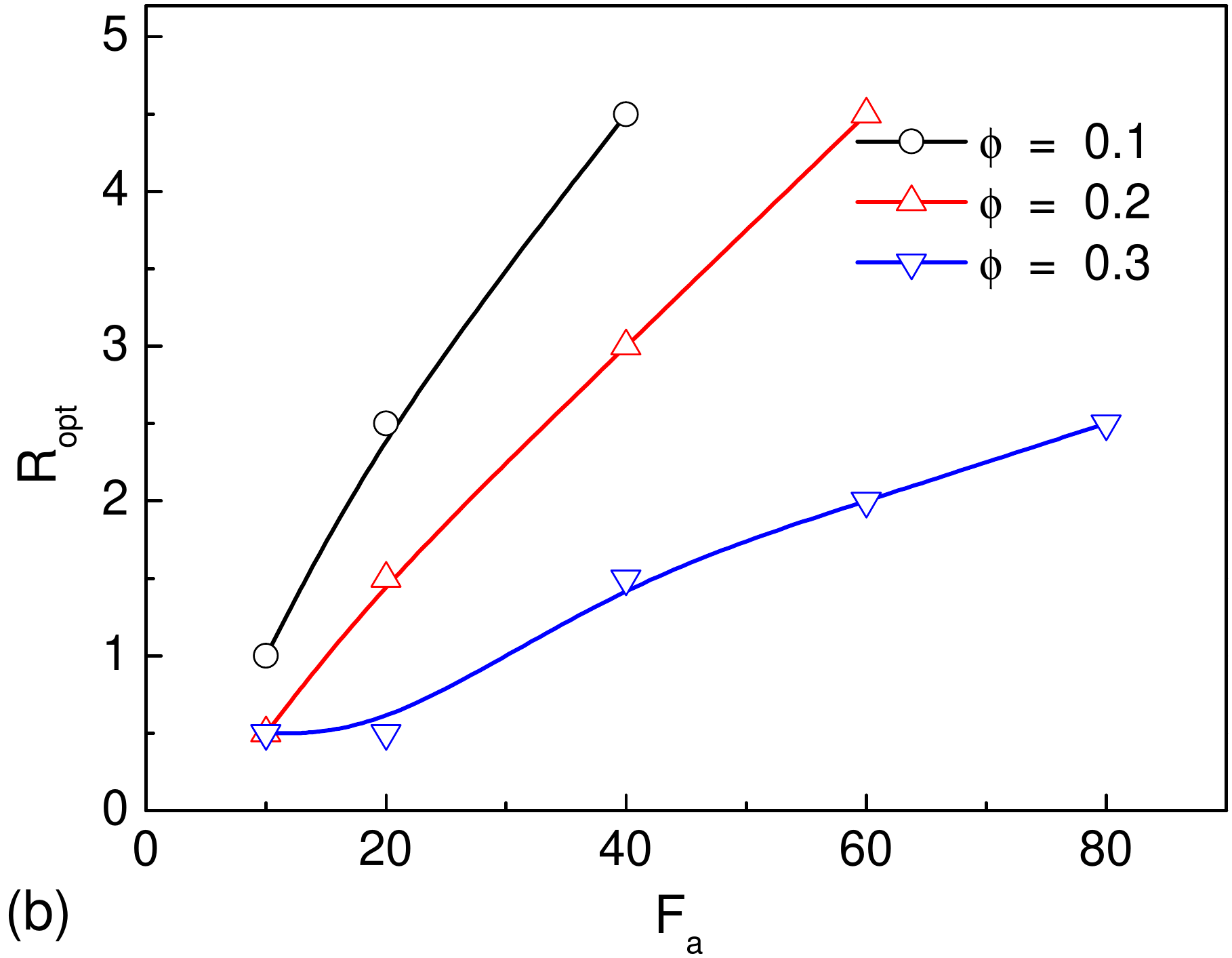}
\par\end{centering}

\textcolor{black}{\caption{\textcolor{black}{(a) The diffusion coefficient $D$ of ABP as a function
of the particle radius $R$, for difference active forces $F_{a}=20,$
40, 60 and 80 and fixed $\phi=0.3$. (b) Dependence of $R_{\text{opt}}$
on active force $F_{a}$ for different number concentrations $\phi=0.1$,
0.2 and 0.3. The solid lines are drawn to guide the eyes.} \label{fig:D_ABP_03}}
}
\end{figure}

\textcolor{black}{We now turn to the ABP. In Fig.\ref{fig:D_ABP_03}(a),
$D$ as a function of $R$ at a fixed concentration $\phi=0.3$ is
presented, for a few different values of propelling force $F_{a}$.
For a relatively small active force, e.g., $F_{a}=20$ as shown in
black line in Fig.\ref{fig:D_ABP_03}(a), $D$ decreases monotonically
with the particle size $R$, which is similar to the case of a passive
particle as shown in Fig.\ref{fig:D_passive}. For larger active forces,
however, $D$ shows an interesting non-monotonic dependence on the
particle size $R$, i.e., $D$ first increases with $R$ until it
reaches a maximum value $D_{max}$ at an optimal size $R_{opt}$ and
then decreases again, as demonstrated clearly in Fig.\ref{fig:D_ABP_03}(a)
for $F_{a}=40,\,60,$ and 80, respectively. With increasing $F_{a}$,
the overall values of $D$ become larger and the optimal size slightly
$R_{opt}$ shifts to larger values. In Fig.\ref{fig:D_ABP_03}(b),
dependence of the optimal particle size $R_{opt}$ on the active force
$F_{a}$ for a few different polymer concentrations $\phi$ are depicted.
For a fixed $\phi$, $R_{opt}$ increases with $F_{a}$ as already
shown in Fig.\ref{fig:D_ABP_03}(a), while $R_{opt}$ decreases with
$\phi$ if $F_{a}$ is fixed. }

The above findings about the non-monotonic dependence of $D$ on $R$
is quite counterintuitive\textcolor{magenta}{{} }\textcolor{black}{at
the first glance,} particularly in terms of the increasing of $D$
with $R$. Generally, one would expect that a larger particle would
diffuse more slowly as the conventional SE relation would tell. For
a passive nano-particle in a polymer solution, although large deviations
from the SE relation were observed and a length-scale viscosity should
be used in replace of the macroscopic viscosity $\eta_{macro}$, $D$
is always a decreasing function of $R$ as already shown in Fig.\ref{fig:D_passive}.
Therefore, the increase of $D$ with $R$ as shown in Fig.\textcolor{blue}{\ref{fig:D_ABP_03}}(a)
must be related to the active feature of the ABP. Indeed, if the active
force is not large enough, as shown for $F_{a}=20$ in Fig.\textcolor{blue}{\ref{fig:D_ABP_03}}\textcolor{black}{(a),
$D$ will still be a decreasing function of $R$, being same to the
case of a passive NP. For large active force, the non-monotonic dependence
of $D$ on $R$ suggests the existence of two competitive factors
that influence the ABP diffusion. }

\subsection{Subdiffusion and Superdiffusion}

In order to understand in more detail about such nontrivial dependence
of $D$ on $R$, we further analyze the short time dynamics of the
ABP by investigating the MSD as a function of time $t$, and compare
it to that of a passive one.

\begin{figure}
\begin{centering}
\includegraphics[width=0.8\columnwidth]{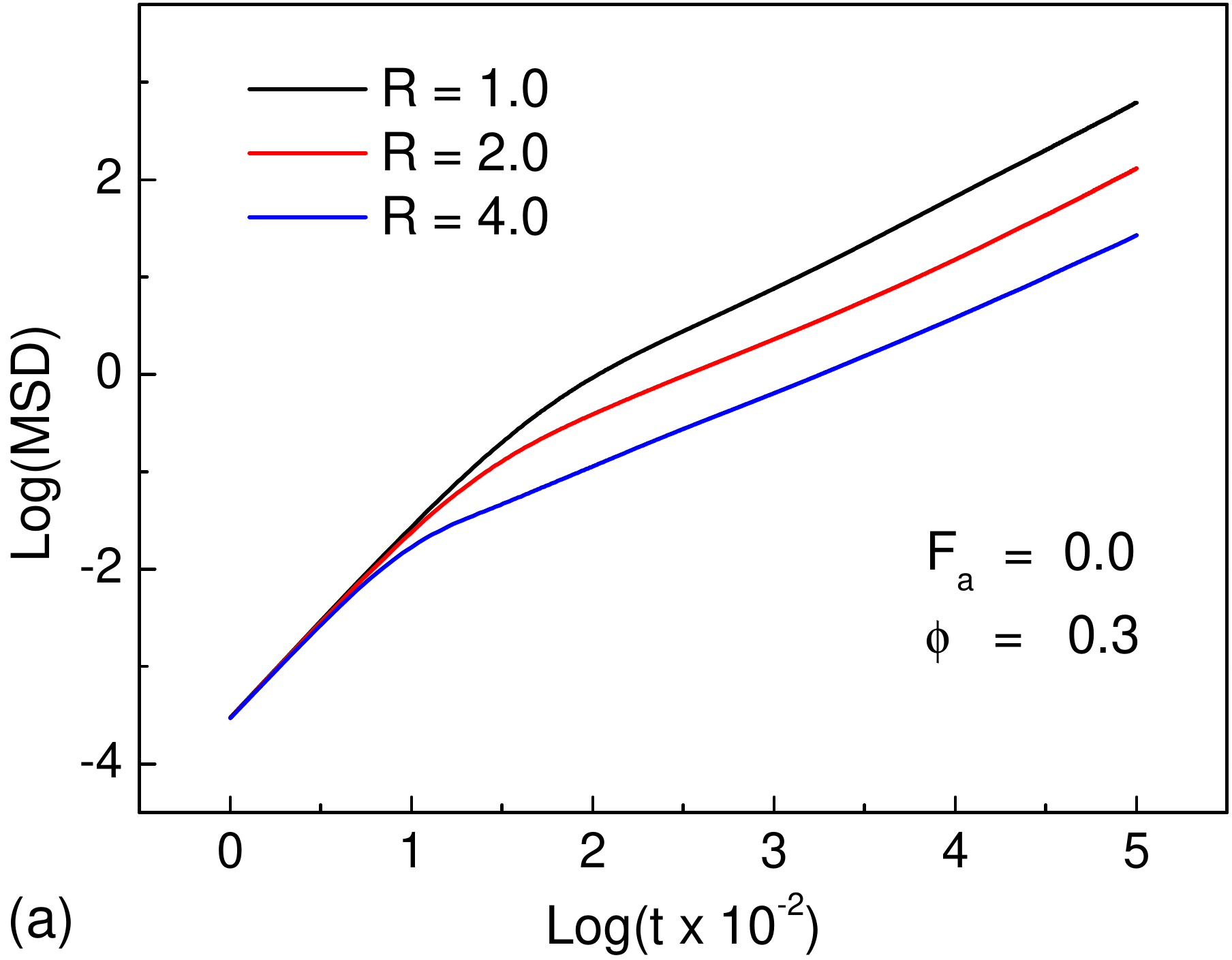}
\par\end{centering}

\begin{centering}
\includegraphics[width=0.8\columnwidth]{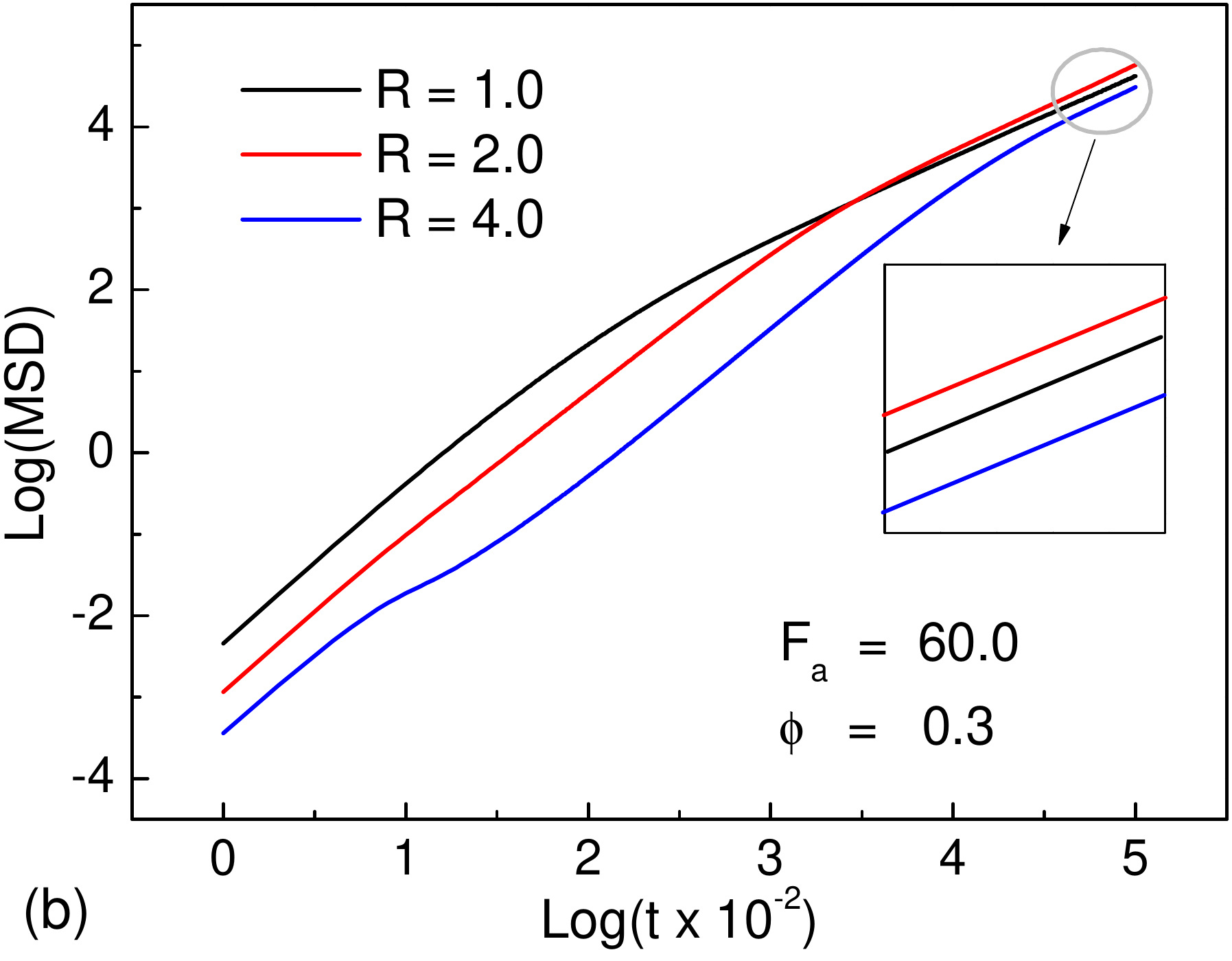}
\par\end{centering}

\caption{MSD of an active NP (a) and a passive NP (b) with size $R=1.0$, 2.0,
4.0. The short line indicates the time scaling at different time scale.
For an active NP, the magnitude of propelling force is fixed at $F_{a}=20.0$.\label{fig:MSD-of-ABP}}
\end{figure}

\begin{figure}
\begin{centering}
\includegraphics[width=0.8\columnwidth]{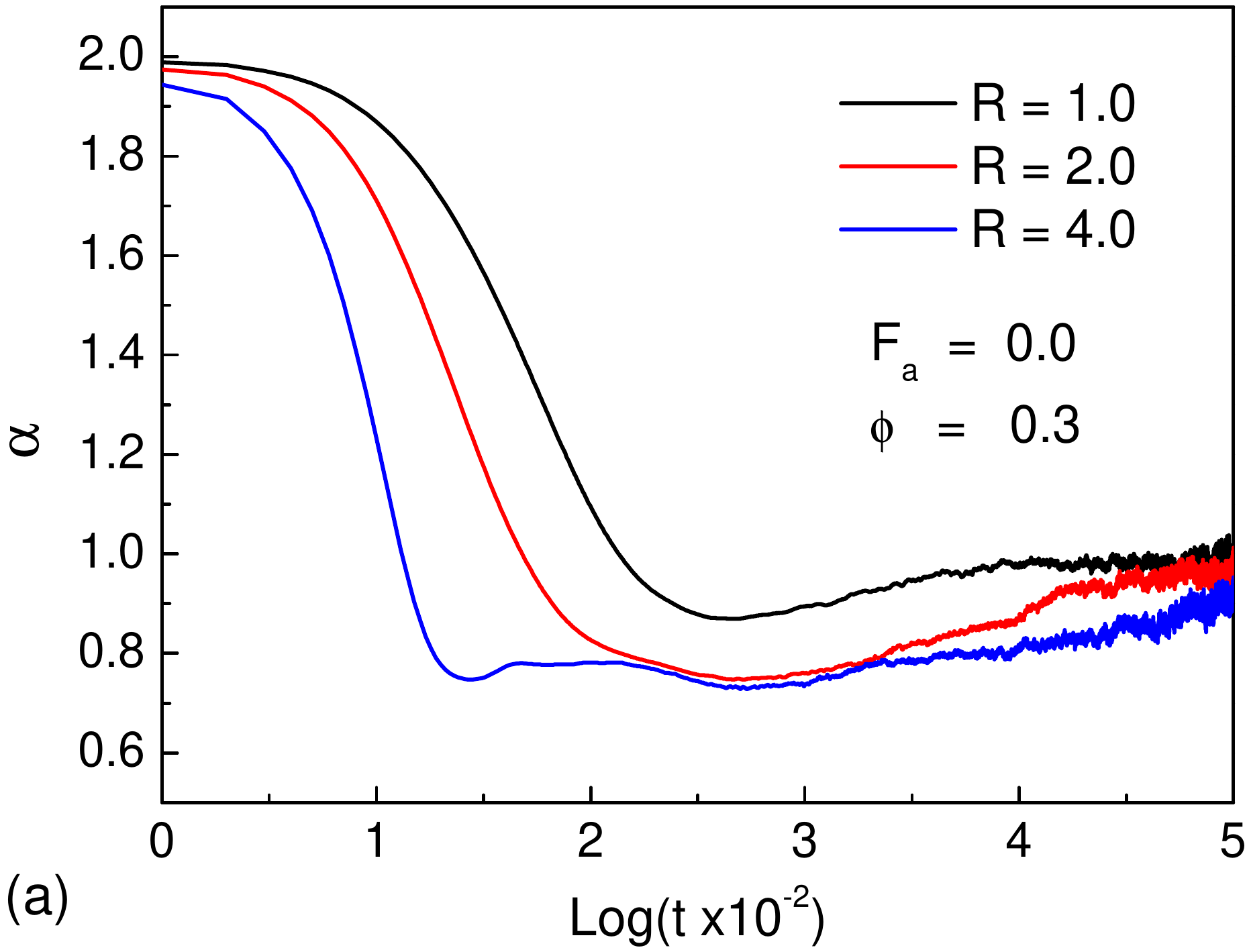}
\par\end{centering}

\begin{centering}
\includegraphics[width=0.8\columnwidth]{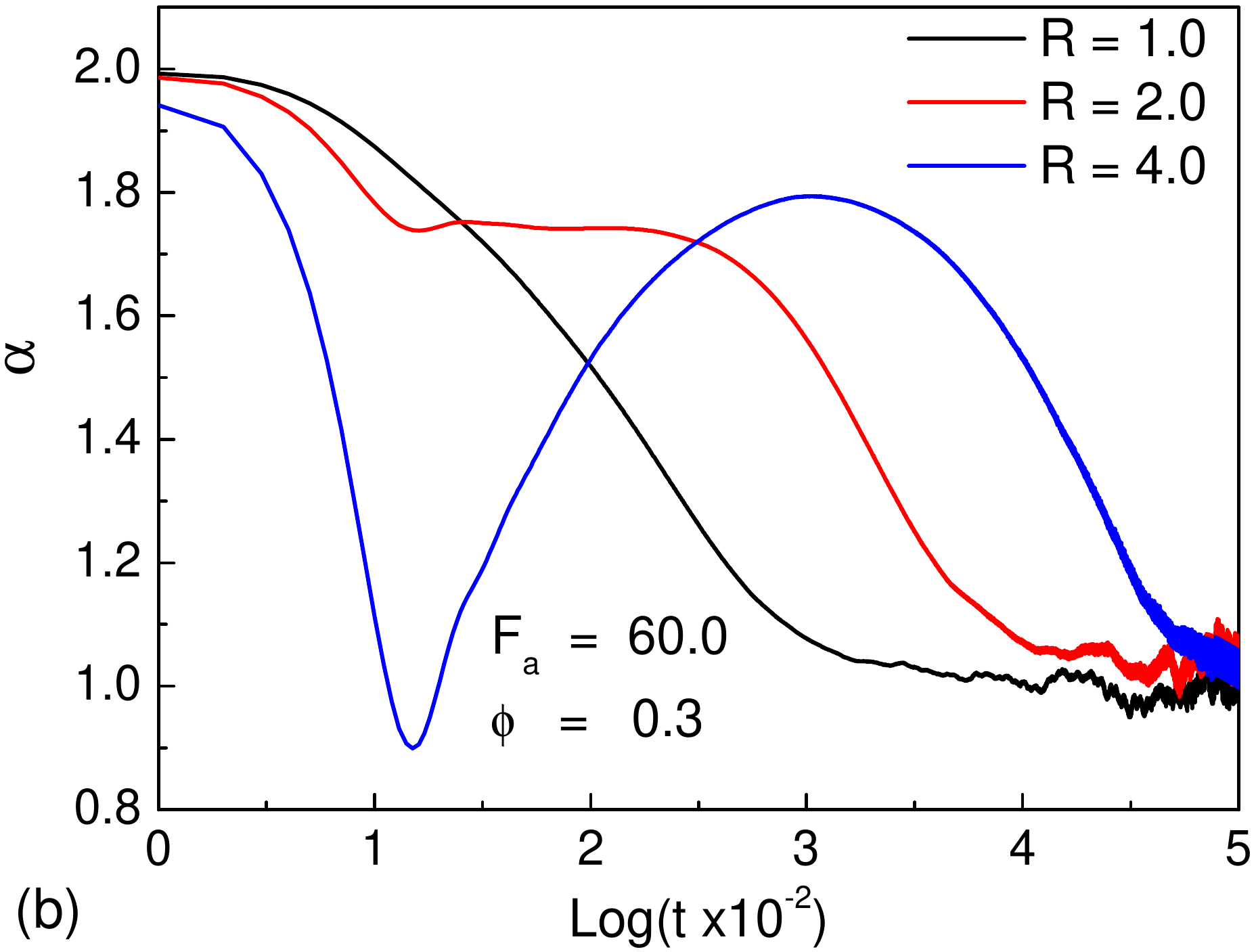}
\par\end{centering}

\caption{The differential of MSD of of a passive NP (a) and an ABP (b) with
size R = 1.0, 2.0 and 4.0 in the polymer solution with concentration
$\phi=0.3$ . $\alpha$ is just the slope of curve in Fig.\ref{fig:MSD-of-ABP}
or the scaling exponent at different time scale. For an ABP depicted
in (b) , the magnitude of propelling force is fixed at $F_{a}=60.0$.
\label{fig:alpha_of_MSD}}
\end{figure}

Fig.\ref{fig:MSD-of-ABP}(a) presents the MSDs for passive NPs with
radius $R=1$, 2 and 4 for $\phi=0.3$. The curves share some common
features, namely, ballistic diffusion at very short time while normal
diffusion at very long time, correspond to $\text{MSD}\sim t^{2}$
and $\text{MSD}\sim t$ respectively. For a small NP, such as $R=1.0$,
the dynamics transformers from ballistic motion to normal diffusion
gradually. With increasing particle size, the NP may experience a
``cage effect'' resulting from the surrounding polymer beads and
the MSD shows a sub-diffusion regime at the middle time scale, wherein
$\text{MSD}\sim t^{\alpha}$ with $\alpha<1$. This is shown more
clearly in Fig.\ref{fig:alpha_of_MSD}(a), where the instantaneous
exponent $\alpha$ is depicted as a function of time corresponding
to Fig.\ref{fig:MSD-of-ABP}(a). Obviously, with increasing the NP
size, the cage effect is more remarkable and $\alpha$ decreases to
a smaller value in the intermediate time scale, corresponding to a
stronger and longer subdiffusion behavior. Also note that the curve
for a larger particle always lies below that for a smaller one in
the whole time range.

For an ABP, however, the behavior is quite different, as shown in
Fig.\ref{fig:MSD-of-ABP}(b) and Fig.\ref{fig:alpha_of_MSD}(b) for
$F_{a}=60$. For this large active force, the diffusion coefficient
$D$ shows non-monotonic dependence on $R$ as demonstrated in last
subsection. For a small particle with $R=1.0$, the behavior shows
no distinct difference from that of a passive one in terms of both
the MSD curve as well as the exponent $\alpha$. For a larger particle
with $R=2.0$, however, we find that the particle undergoes a much
longer superdiffusion time regime with $1<\alpha<2$ before it finally
reaches the long time normal diffusion regime. Interestingly, the
exponent $\alpha$ shows an apparent plateau at $\alpha\sim1.7$ as
shown in Fig.\ref{fig:alpha_of_MSD}(b) which spans two orders of
magnitude of time. Such a superdiffusion behavior with $\alpha$ slightl\textcolor{black}{y
smaller t}han 2 implicates that the particle moves more persistently
along a direction than randomly along different directions. Note that
this persistence of motion along a direction reflects the very feature
of an active Brownian particle. For an even larger particle $R=4.0$,
one can see that the exponent $\alpha$ first decreases sharply to
a value at 0.9 in the short time range, namely indicated a sub-diffusion
behavior, and then increases again to a high value close to 2 in the
intermediate range before it finally decreases to the normal diffusion
value $\alpha=1$. The sharp decrease of $\alpha$ in the short time
range should result from the cage effect of the surrounding polymers.
Due to the large activity of the particle, however, such a cage effect
can not last for a long time and finally the particle jumps out of
this cage leading to the increase of $\alpha$ in the intermediate
time range. In the short time range, ABP with a smaller size moves
faster than a larger one as shown in Fig.\ref{fig:MSD-of-ABP}(b).
However, in the long time limit, the curve for a middle particle size
$R=2$ lies above those of $R=1$ and 4, corresponding to a maximum
value of long time diffusion coefficient $D$ for $R=2$ compared
to those others two in accordance with Fig.\ref{fig:D_ABP_03}.

The above analysis suggests that the occurrence of an optimal particle
size for ABP diffusion in polymer solution is the consequence of two
competitive effects. One is that the cage effect of the surrounding
polymers which would become stronger as the particle size becomes
larger. Without activity, this cage effect would lead to subdiffusion
behavior of a particle and decrease of the long time diffusion coefficient
$D$. Such a cage effect results in a length-scale dependent viscosity
experienced by the particle as described in Fig.\ref{fig:D_passive}(b)
which increases with $R$. The other is the persistence motion due
to the particle activity which would become longer as the particle
size increases. As well known for an ABP and described in the model
section, the persistence time $\tau_{p}$ for an isolated ABP is given
by $\left(2D_{r}\right)^{-1}$, where $D_{r}$ scales as $R^{-2}$
for the ABP as shown in section \ref{sec:Simulation-method}. Therefore,
the ABP would move along the propelled direction longer as the particle
size gets larger, leading to superdiffusive behavior and thus accelerating
the long time diffusion. When the particle size is small, e.g. $R=1$,
both cage and persistence effects are not significant and the particle
transfers gradually from ballistic to diffusion motion for both passive
and active particles. For a relatively larger particle, the persistence
effect would dominate, thus leading to increase of $D$. If the particle
size is too large, however, the cage effect would dominate and the
diffusion coefficient would decrease again. Beside\textcolor{magenta}{s},
increasing the magnitude of $F_{a}$, the persistence motion would
be enhanced. Clearly, this enhancement will be more apparent for a
bigger ABP with longer $\tau_{p}$. Therefore, the stronger $F_{a}$
promotes $R_{opt}$ to a bigger value as shown in Fig.\ref{fig:D_ABP_03}(b).

\subsection{Effective Viscosity $\eta_{\text{eff}}$}

\textcolor{black}{As discussed in Section 3.1, a passive particle
would experience an effective viscosity $\eta_{\text{eff}}$ that
is dependent on its size $R$, which could be much smaller than the
macroscopic zero-shear viscosity $\eta_{\text{macro}}$. For a passive
particle, this effective viscosity is determined according to the
SE relation, $D=k_{\text{B}}T/\left(6\pi\eta_{\text{eff}}R\right)$.
It is thus also interesting for us to ask the question what is the
effective viscosity the ABP experiences in the polymer solution. }

\textcolor{black}{At first thought, one may also just use the SE relation
to obtain this effective viscosity $\eta_{\text{eff}}$, i.e., $\eta_{\text{eff}}=k_{\text{B}}T/\left(6\pi DR\right)$
where $D$ is the long time translational diffusion coefficient obtained
by simulation above. Nevertheless, this may not be appropriate for
an active particle, since we must take into account the particle activity
which would lead to an active contribution to $D$. As already discussed
in the model section, for an isolated ABP of radius $R$ in a simple
fluid with friction coefficient $\gamma$, the dynamics can be described
by the following overdamped Langevin equation (LE),
\begin{align}
\frac{d\mathbf{r}}{dt} & =\gamma^{-1}F_{a}\textbf{n}+\sqrt{2D_{t}}\mathbf{\bm{\xi}}\left(t\right)\label{eq:overdamped_LE}\\
\frac{d\mathbf{n}}{dt} & =\bm{\zeta}\left(t\right)\times\mathbf{n}\nonumber
\end{align}
where $\bm{\xi}\left(t\right)$ and $\bm{\zeta}\left(t\right)$ are
both Gaussian white noise vectors with zero mean and unit(tensor)
variance, $D_{t}=k_{B}T/\gamma$ and $D_{r}=3D_{t}/4R^{2}$. In a
coarse-grained time scale, it was shown that the self-propulsion force
for ABP can be mapped into a colored noise\citep{bechinger2016active},
i.e.,
\[
\left\langle \mathbf{n}\left(t\right)\mathbf{n}\left(t'\right)\right\rangle \simeq\frac{1}{3}e^{-\left|t-t'\right|/\tau_{p}}\mathbf{1}
\]
where $\mathbf{1}$ denotes the unit tensor and $\tau_{p}=\left(2D_{r}\right)^{-1}$
denotes the persistence time of the self-propulsion force. With this
approximation, one can obtain the mean-square displacement (MSD) of
the ABP as follows, }

\[
\left\langle \delta r^{2}\left(t\right)\right\rangle =\frac{2F_{a}^{2}\tau_{p}}{\gamma^{2}}\left[t+\tau_{p}\left(e^{-t/\tau_{p}}-1\right)\right]+6D_{t}t
\]
In the long time limit, this gives the diffusion coefficient
\[
D=\lim_{t\to\infty}\frac{1}{6}\frac{\left\langle \delta r^{2}\left(t\right)\right\rangle }{t}=D_{t}+\frac{F_{a}^{2}\tau_{p}}{3\gamma^{2}}
\]
If we use the fact that $\tau_{p}=\left(2D_{r}\right)^{-1}$ and $D_{r}=3D_{t}/4R^{2}$,
then we have
\begin{equation}
D=\frac{k_{B}T}{\gamma}+\frac{2F_{a}^{2}R^{2}}{9k_{B}T\gamma}=\frac{1}{6\pi\eta_{0}R}\left(k_{\text{B}}T+\frac{2F_{a}^{2}R^{2}}{9k_{\text{B}}T}\right)\label{eq:D_Simple}
\end{equation}
where we have used the Stokes relation $\gamma=6\pi\eta_{0}R$ in
the simple fluid with $\eta_{0}$ the zero-shear viscosity. This analysis
shows how the long time diffusion coefficient $D$ depends on the
viscosity $\eta_{0}$ of a simple fluid. The first term is surely
the SE relation, while the second term denotes the contribution from
particle activity. For small $T$ or large $F_{a}$, the second active
term would dominate and the relation between $D$ and $\eta_{0}$
is totally different from the SE relation.

Actually, eqn (\ref{eq:D_Simple}) provides us a scheme to introduce
an effective viscosity experienced by the ABP in the surrounding solution.
We consider now that the ABP is moving in a pure fluid with effective
viscosity $\eta_{\text{eff}}$, whose dynamics is also described by
eqn (\ref{eq:overdamped_LE}), but with $\gamma$ replaced by an effective
$\gamma_{\text{eff}}^{a}=6\pi\eta_{\text{eff}}^{a}R$ (the superscript
`a' stands for `active'). The (short time) translational diffusion
constant $D_{t}$ is determined then by $\gamma_{\text{eff}}^{a}$
through fluctuation-dissipation theorem $D_{t}=k_{\text{B}}T/\gamma_{\text{eff}}^{a}$,
which in turn gives $D_{r}$. Clearly, the long time diffusion coefficient
$D$ would also be given by eqn (\ref{eq:D_Simple}) with $\eta_{0}$
replaced by $\eta_{\text{eff}}^{a}$, i.e.,
\begin{equation}
D=\frac{1}{6\pi\eta_{\text{eff}}^{a}R}\left(k_{\text{B}}T+\frac{2F_{a}^{2}R^{2}}{9k_{\text{B}}T}\right)\label{eq:D_Eta_Eff}
\end{equation}
 Therefore, the effective viscosity of the polymer solution experienced
by the ABP could be defined as
\begin{equation}
\eta_{\text{eff}}^{a}=\frac{1}{6\pi DR}\left(k_{\text{B}}T+\frac{2F_{a}^{2}R^{2}}{9k_{\text{B}}T}\right)\label{eq:Eta_eff}
\end{equation}

\begin{figure}
\begin{centering}
\includegraphics[width=0.8\columnwidth]{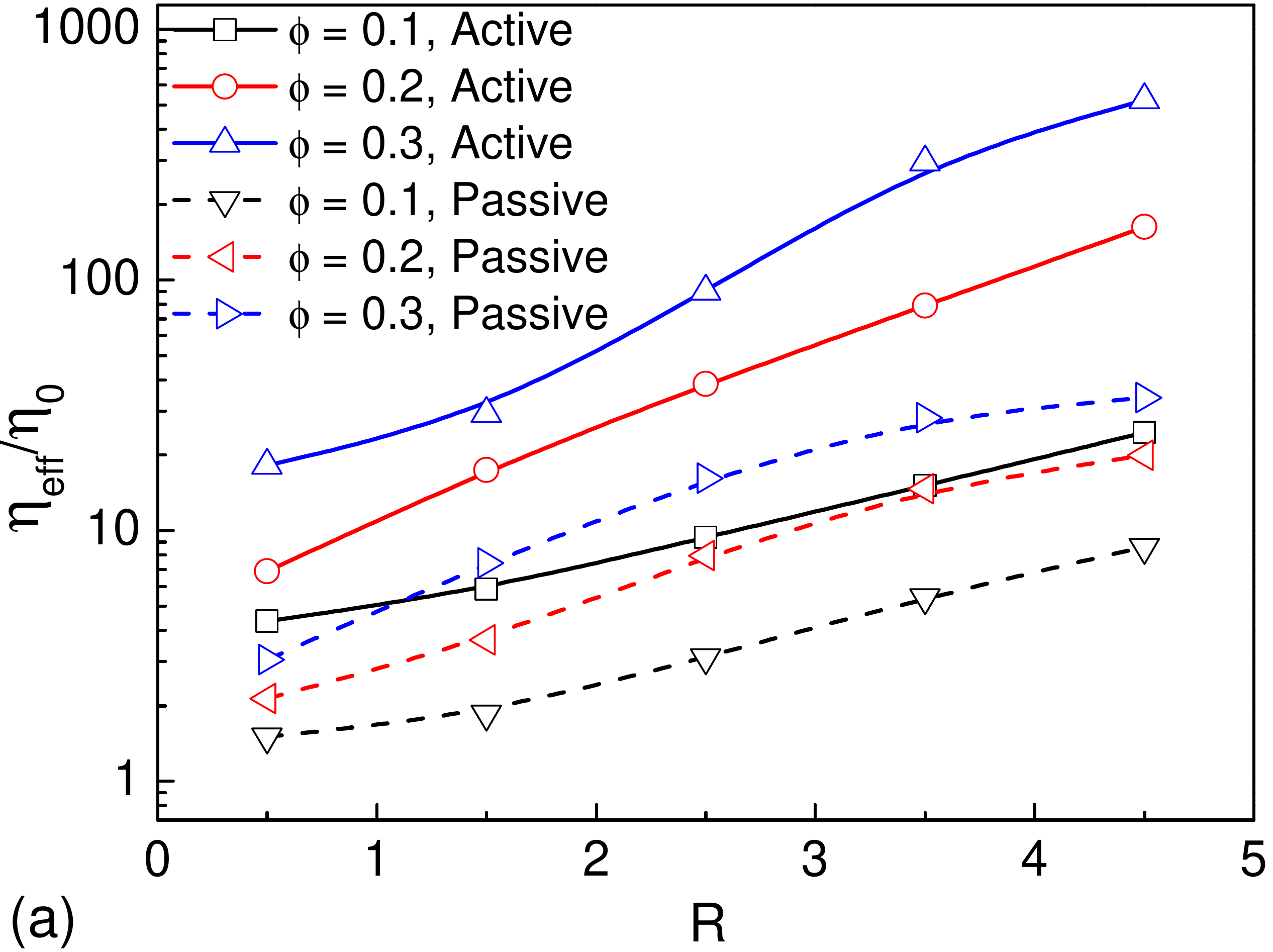}
\par\end{centering}

\begin{centering}
\includegraphics[width=0.8\columnwidth]{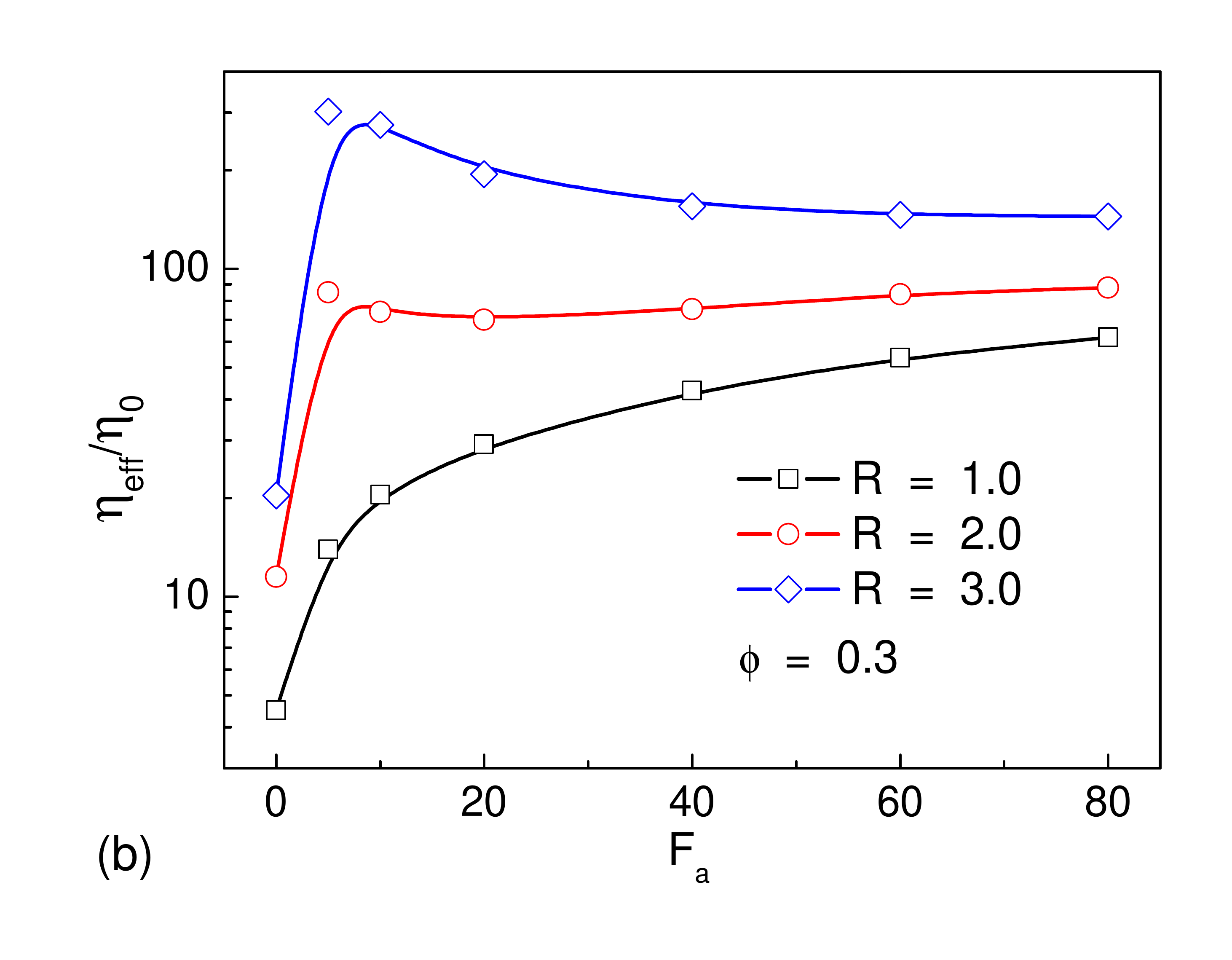}\caption{Effective viscosity $\eta_{\text{eff}}$ as a function of (a) particle
size $R$ for different values of concentration $\phi=0.1$, 0.2 and
0.3 with fixed active force $F_{a}=20$, and (b) active force $F_{a}$
for different particle size $R=1.0$, 2.0 and 3.0 with fixed concentration
$\phi=0.3$. \textcolor{black}{Note that $\eta_{\text{eff}}$ is reduced
by the viscosity of pure solution $\eta_{0}$}\label{fig:Effective-viscosity}}

\par\end{centering}

\end{figure}

In Fig.\ref{fig:Effective-viscosity}(a), we show the effective viscosity
$\eta_{\text{eff}}^{a}$ calculated by eqn (\ref{eq:Eta_eff}) as
functions of the particle size $R$ for different concentrations $\phi=0.1$,
0.2 and 0.3, where the values of $D$ are obtained by simulations
as shown in Fig.\ref{fig:D_ABP_03}. The amplitude of the active force
is $F_{a}=20$. Also shown are the values for a passive particle,
which were already presented in Fig.\ref{fig:D_passive}(b). As can
be seen, $\eta_{\text{eff}}^{a}$ increases with $R$ as expected,
similar to the case of $\eta_{\text{eff}}$ for a passive particle.
Interestingly, $\eta_{\text{eff}}^{a}$ is much larger than $\eta_{\text{eff}}$
\textcolor{black}{as shown in Fig.\ref{fig:Effective-viscosity}(a}\textcolor{blue}{),}
indicating that the active particle seems to move in a ``more viscous''
fluid than the passive one. Nevertheless, one should be careful to
draw the conclusion that particle activity induces thickening of the
polymer solution, since $\eta_{\text{eff}}^{a}$ here is defined via
eqn (\ref{fig:Effective-viscosity}). As discussed above, this equation
is obtained by modeling the motion of ABP in polymer solution as if
it were in an viscous fluid with $\eta_{\text{eff}}^{a}$, while keeping
fluctuation-dissipation theorem and all other features of ABP unchanged.
Indeed, the effective viscosity experienced by an ABP characterize
more the local environment surrounding the particle, and it is not
identical to the zero-shear macro-viscosity of the solution. Our arguments
here indicate that the ABP does feel a much more viscoelastic local
environment than a passive one, otherwise it would diffuse much faster.
In Fig.\ref{fig:Effective-viscosity}(b), the effective viscosity
$\eta_{\text{eff}}^{a}$ as a function of active force amplitude $F_{a}$
for different particle size $R$ are presented. If the particle size
is relatively small, say $R=1.0$, $\eta_{\text{eff}}^{a}$ is a monotonic
increasing function of $F_{a}$, implying that a more active particle
feels a more viscoelastic local fluid. For large particle sizes, however,
more interesting features can be observed: $\eta_{\text{eff}}^{a}$
shows a non-monotonic dependence on $F_{a}$. For $R=3.0,$ for instance,
$\eta_{\text{eff}}^{a}$ first increases sharply to a very large value
with increment of $F_{a}$ and then decreases relatively slowly to
a moderate value when $F_{a}$ is large. Since $\eta_{\text{eff}}^{a}$
is calculated via definition in our present work, the mechanism behind
these interesting observations is still open to us and may deserve
more detailed study in future works.

\subsection{Scaling of $D$ with active force $F_{a}$}

Another feature shown in Fig.\ref{fig:Effective-viscosity}(b) is
that $\eta_{\text{eff}}^{a}$ becomes not sensitive to $F_{a}$ if
$F_{a}$ is large enough. Now considering eqn (\ref{eq:D_Eta_Eff})
for the long time diffusion constant $D$, one can see that the second
term would dominate if $F_{a}$ is large. Therefore, $D$ would scale
approximately as $F_{a}^{2}$ in the range of large $F_{a}$. Motivated
by this observation, we have also investigated quantitatively how
$D$ depends on $F_{a}$ in the whole range.

\begin{figure}
\begin{centering}
\includegraphics[width=0.8\columnwidth]{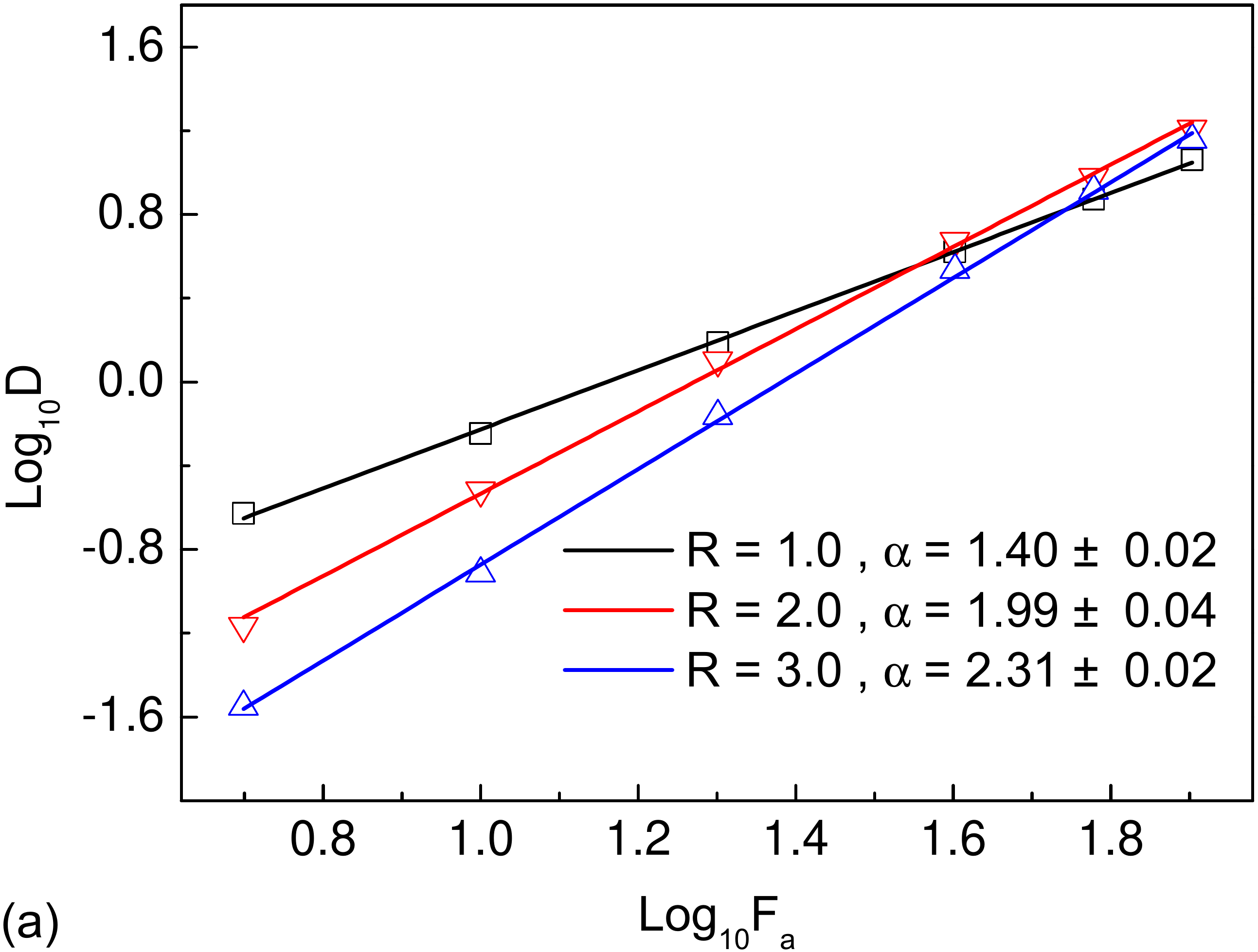}
\par\end{centering}

\begin{centering}
\includegraphics[width=0.8\columnwidth]{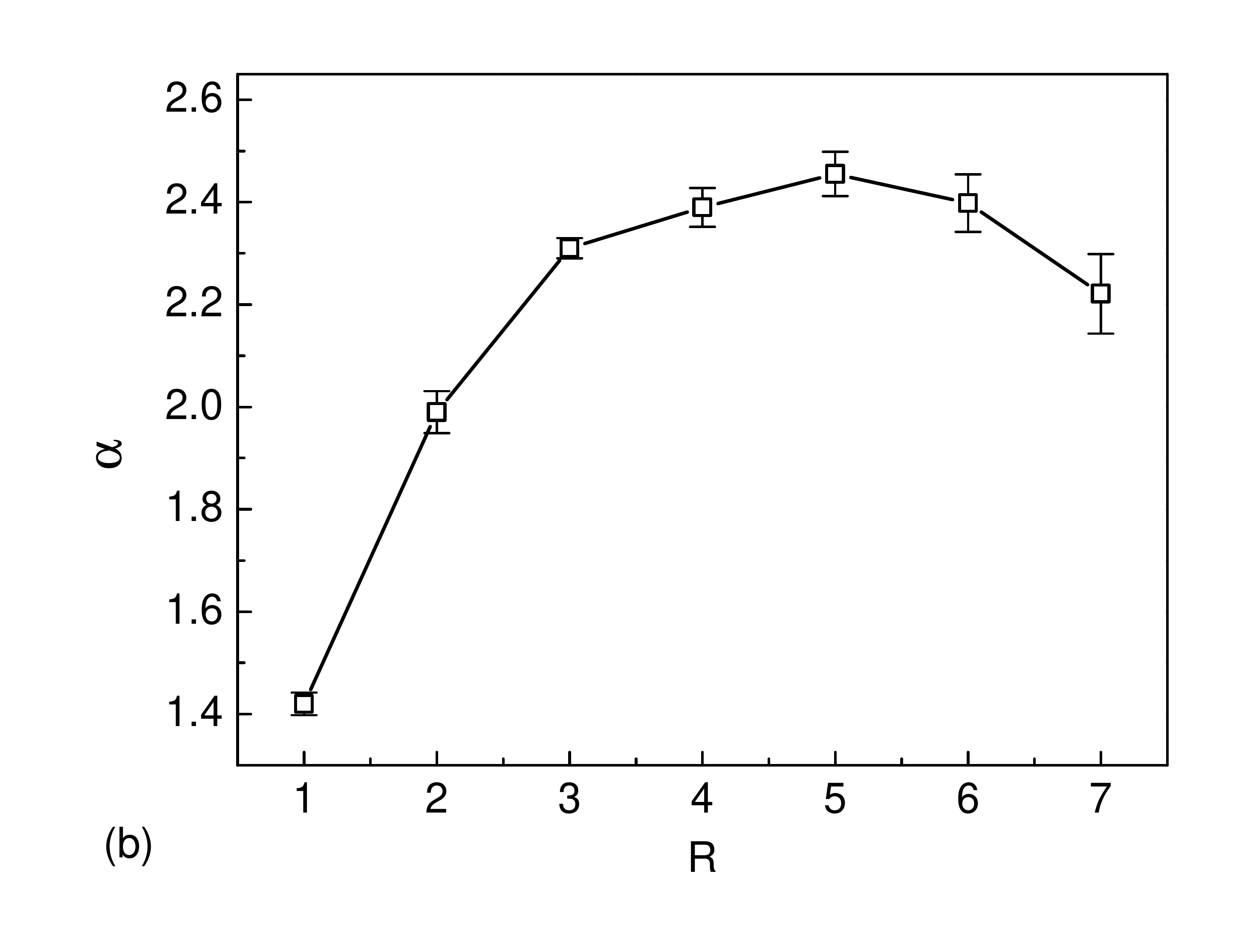}
\par\end{centering}

\caption{(a)dependence of diffusion coefficient $D$ on propelling force $F_{a}$
in a scaling form. The radius of NP are fixed at $R=1.0,$ 2.0 and
3.0, represented by black, red and blue dots, respectively. The solid
lines are scaling fits to the data(b)The scaling factor of the function
between $D$ and $F_{a}$ in a fixed size of NP. The solid line is
drawn to guide eyes. \label{fig:scaling }}
\end{figure}

Surprisingly, we find that $D$ shows a rather good power-law scaling
with $F_{a}$, i.e., $D\sim F_{a}^{\alpha}$, as shown in Fig.\ref{fig:scaling }(a)
for different particle sizes and fixed concentration. For $R=1.0$,
2.0 and 3.0, the scaling exponent $\alpha$ reads approximately 1.40,
1.99 and 2.31, respectively. In Fig.\ref{fig:scaling }(b), the dependence
of $\alpha$ exponent on particle size $R$ is depicted, where one
observes a rather interesting non-monotonic variation: $\alpha$ increases
from 1.4 at $R=1.0$ to a maximum value $\alpha\simeq2.4$ at $R=5.0$
and then decreases again to about 2.2 at $R=7.0$ which is the largest
particle size considered in the present work (to get a reliable data
at $R>$5.0 that avoiding the finite-size effects, we have to extend
the system to $l=30.0\sigma_{0}$ consisting of totally 126 polymers).
In the limit of very large particle size, one would imagine that the
polymer solution can be viewed as a simple fluid, and the exponent
would become 2.0 again.

In the current stage, we are yet not able to understand the power-law
scaling between $D$ and the active force $F_{a}$. eqn (\ref{eq:D_Eta_Eff})
would simply give $D\sim F_{a}^{2}R/\eta_{\text{eff}}^{a}$ when the
active term dominates, which seems to suggest an exponent to be 2.0
if $\eta_{\text{eff}}^{a}$ is not dependent on $F_{a}$. Nevertheless,
eqn (\ref{eq:D_Eta_Eff}) is just a definition of $\eta_{\text{eff}}^{a}$
which is obtained from the simulated value of $D$, and the calculated
values of $\eta_{\text{ef}\text{f}}^{a}$ do depend on $F_{a}$ as
already shown in Fig. \ref{fig:Effective-viscosity}. Therefore, to
understand the power-law dependence of $D$ on $F_{a}$, one has to
derive a separate theory for the diffusion of ABP in polymer solutions,
which is important but beyond the scope of current study.

Here in the present paper, we would like to take a qualitative description
to highlight the active effects. According to the MCT framework to
study the diffusion of a passive NP in polymer solution, the long
time diffusion coefficient can be decomposed into two different parts,
i.e., $D=D_{\text{micro}}+D_{\text{hydro}}$, where $D_{\text{hydro}}$
is the conventional SE term, while $D_{\text{micro}}$ is due to the
microscopic level interactions between the particle and polymer molecules.
The friction related to $D_{\text{micro}}$ results from direct\textit{
binary collisions} between the NP and the polymer beads and \textit{density
fluctuation }of the solution. For large particles in entangled solutions
with strong topological constraints, however, one should take into
account another possible mechanism for diffusion, namely, the \textit{hopping}
process\citep{cai2015hopping}, wherein the particle can diffuse by
overcoming barrier between neighboring confinement cells. Now we take
the particle activity into account. Clearly, enhancement of particle
activity would lead to more frequent direct collisions between the
particle and polymer beads, thus leading to a larger effective viscosity
compare to that experienced by a passive particle. Since we only consider
a single particle here, the density fluctuation contribution to the
friction would not change much with the variation of the particle
activity. Nevertheless, particle activity would facilitate the hopping
process if the particle is large enough, which would result in a relatively
smaller effective viscosity. Therefore, for a small particle such
as $R=1.0$, the effective viscosity experienced by an ABP would increase
with the amplitude $F_{a}$, since hopping diffusion is not relevant
for small particles. In this case, the scaling exponent $\alpha$
would be less than $2.0$ since $\eta_{\text{eff}}$ scale as $F_{a}^{\lambda}$
with $\lambda>0$. For a large particle such as $R=3.0$, both binary
collisions and hopping processes will be enhanced by the particle
activity. These two effects competes with each other, leading to the
non-monotonic dependence of $\eta_{\text{eff}}^{a}$ with $F_{a}$
as demonstrated in Fig.\ref{fig:Effective-viscosity}(b). In the small
$F_{a}$ range, enhanced binary collision dominates leading to an
sharp increase of $\eta_{\text{eff}}^{a}$, while for large $F_{a}$,
enhanced hopping dominates which results in decrease of $\eta_{\text{eff}}^{a}$.
In this case, the exponent $\alpha$ should be larger than 2.0 in
the large $F_{a}$ range. For a particle of proper size, the positive
and negative effects of particle activity may cancel, which leading
to weak dependence of $\eta_{\text{eff}}^{a}$ on $F_{a}$ and the
exponent $\alpha$ would be approximately 2.0. For a very large particle
size, however, the effect that activity enhances hopping may become
weaker than that for a relatively smaller particle, such that $\alpha$
would decrease with $R$. All these features are consistent with the
observations in Fig.\ref{fig:scaling }(b).

\section{\label{sec:Conclusion}Conclusion}

In conclusion, we have used Langevin Dynamic simulation method to
investigate the diffusion behavior of an ABP in semidilute polymer
solution. Extensive simulations indicate that the activity can markedly
enhance the diffusion of NP in this complex environment, respected
to a passive NP. Interestingly, we have found that the dependence
of long time diffusion coefficient $D$ on the particle radius $R$
is non-monotonic and $D$ reaches a maximum value at a certain optimal
value $R_{opt}$, if exerted a propelling force $F_{a}$ strong enough.
Consequently, the NP with a bigger size would get a higher $D$ than
a relatively smaller one, which is greatly against the common sense
that a larger particle usually diffuse slower for a passive NP.

The subsequent analysis from the short time dynamics reveals that
this abnormal phenomenon is due to the two fold impacts of ABP on
diffusion. In detail, the increasing size of ABP gains more obstruction
from the polymer beads that leads to more apparent cage effects, which
finally slows down the diffusion. However, on the other side, a bigger
size could usually make the rotation tougher and result in the longer
rotational relaxation time $\tau_{p}$. This would lead to longer
persistence motion along a direction and cause superdiffusion behavior
which would cover the subdiffusion from the cage effect, and finally
facilitates the diffusion. It is the competition between the persistence
motion and the cage effect that leads to the non-monotonic dependence
of the long time diffusion $D$ on the particle size $R$.

We have also introduced a phenomenological model to describe the ABP
dynamics, assuming that the ABP is moving in a simple viscous fluid
with effective viscosity $\eta_{\text{eff}}^{a}$. We find that this
effective viscosity shows strongly dependency on the particle size
$R$ as well as active force $F_{a}$. Interestingly, this effective
viscosity experienced by the ABP is larger than that experience by
a passive nano-particle of the same size, which means that the ABP
feels a much more viscoelastic local environment, otherwise it would
diffuse much faster. For an ABP of small size, $\eta_{\text{eff}}^{a}$
increases monotonically with $F_{a}$, while for a large ABP, $\eta_{\text{eff}}^{a}$
can even show a non-monotonic dependence on $F_{a}$ bypassing a maximum.

A more striking finding is that $D$ shows a power-scaling with the
active force $F_{a}$ in an excellent manner. The exponent $\alpha$
is not equal to the value 2.0 that observed in a simple liquid, but
non-monotonically depends on $R$. It increases from a value quite
smaller than 2.0 to a maximum value at about 2.5, and then decreases
again to 2.0 if $R$ is large enough. Although a rigorous theory for
the diffusion behavior of ABP in polymer solutions is not available
at the current stage, we have tried to understand the effect of activity
on particle diffusion through its influence on the binary collisions
between the particle and polymer beads and on the hopping process
of a large particle out of confinement cells.

Our results indicate that activity combined with the inner structure
of the polymer solution indeed largely affects the diffusion dynamics
of an active particle, both short time and long time, which is greatly
different to a passive one. We believe that our work can open more
perspectives on the study of active matter in complex solutions and
may shed some new lights on understanding such an important process
in real biological systems.
\begin{acknowledgments}
This work is supported by the Ministry of Science and Technology of
China(Grant Nos. 2013CB834606, 2016YFA0400904), by National Science
Foundation of China (Grant Nos. 21673212, 21521001, 21473165, 21403204),
and by the Fundamental Research Funds for the Central Universities
(Grant Nos. WK2030020028, 2340000074).
\end{acknowledgments}

%

\end{document}